\documentclass[twocolumn,showpacs,preprintnumbers,amsmath,amssymb]{revtex4}

\usepackage{epsf}
\usepackage{graphicx}
\usepackage{dcolumn}
\usepackage{bm}

\begin{document}


\title{Observational Constraints on Visser's Cosmological Model}

\author{M.E.S.Alves $^1$}
 \email{alvesmes@das.inpe.br}
\author{F.C. Carvalho $^{1,2}$}
 \email{fabiocc@das.inpe.br}
\author{J.C.N.de Araujo $^1$}
 \email{jcarlos@das.inpe.br}
\author{O.D.Miranda $^1$}%
 \email{oswaldo@das.inpe.br}
\author{C.A. Wuensche $^1$}
 \email{alex@das.inpe.br}
\author{E.M. Santos $^3$}
 \email{emoura@if.ufrj.br
}
\affiliation{ $^1$INPE - Instituto Nacional de Pesquisas
Espaciais - Divis\~ao de Astrof\'isica,
Av.dos Astronautas 1758, S\~ao Jos\'e dos Campos, 12227-010 SP, Brazil\\
$^2$ UERN - Universidade do Estado do Rio Grande do Norte,
Mossor\'o, 59610-210, RN, Brazil \\
$^3$ UFRJ - Universidade Federal do Rio de Janeiro, Rio de
Janeiro, 21945-970, RJ, Brazil}



\begin{abstract}
Theories of gravity for which gravitons can be treated as massive
particles have presently been studied as realistic modifications
of General Relativity, and can be tested with cosmological
observations. In this work, we study the ability of a recently
proposed theory with massive gravitons, the so-called Visser
theory, to explain the measurements of luminosity distance from
the Union2 compilation, the most recent Type-Ia Supernovae (SNe
Ia) dataset, adopting the current ratio of the total density of
non-relativistic matter to the critical density ($\Omega_m$) as a
free parameter. We also combine the SNe Ia data with constraints
from Baryon Acoustic Oscillations ({\rm BAO}) and {\rm CMB}
measurements. We find that, for the allowed interval of values for
$\Omega_m$, a model based on Visser's theory can produce an
accelerated expansion period without any dark energy component,
but the combined analysis (SNe Ia$+$BAO$+$CMB) shows that the
model is disfavored when compared with $\Lambda$CDM model.
\end{abstract}


\pacs{98.80.-k, 95.36.+x, 95.30.Sf}
\maketitle


\section{\label{sec:intr}Introduction}


The current Universe's energy budget is a consequence of the
convergence of independent observational results that led to the
following distribution of the energy densities of the Universe:
4\% for baryonic matter, 23\% for dark matter and 73\% for dark
energy \cite{spergel06}. The key observational results that
support this picture are: mesurements of luminosity distance as a
function of redshift for distant supernovae
\cite{ApJ_517_565,AstronJ116_1009,ApJ659_98}, anisotropies in the
Cosmic Microwave Background (CMB) observed by the WMAP satellite
\cite{0803.0547} and the Large Scale Structure (LSS) matter power
spectrum inferred from galaxy redshift surveys such as the Sloan
Digital Sky Survey (SDSS) \cite{PRD69_103501} and 2dF Galaxy
Redshift Survey (2dFGRS) \cite{MNRAS362_505}.

In order to explain all the currently available cosmological data,
the cosmological concordance model $\Lambda$CDM need to appeal to
two exotic components, the so called dark matter and dark energy.
The latter drives the late time accelerated expansion of the
Universe and it is one of the greatest challenges for the current
cosmology. Indeed, the physical nature of the dark energy is a
particularly complicated issue to address in the $\Lambda$CDM
context, due to its unusual properties. It behaves as a
negative-pressure ideal fluid smoothly distributed through space.
One can ask if the accelerating expansion of the Universe might
indicate that Einstein's theory of gravity is incomplete, i.e.,
can an alternative theory of gravity explain consistently the
late-time cosmic acceleration without recurring to dark energy?

There are several alternative approaches based on the idea of
modifying gravity. Currently, one of the most studied alternative
gravity theories is the so called $f(R)$ gravity, whose basic idea
is to add terms which are powers of the Ricci scalar $R$ to the
Einstein-Hilbert Lagrangian \cite{PRD70_043528,GRG36_1765,
PLB669_14, JCAP0809_008, PRD76_083513,Alves2009}.

Recently, M. Visser proposed a modification of the general
relativity (GR) where the gravitons can be massive particles
\cite{vis1998}. In particular, several authors have studied the
limits that can be imposed to the graviton mass using different
approaches. For example, from analysis of the planetary motions in
the solar system it was found that $m_g < 7.8 \times 10^{-55}$g
\cite{Tal1988}. Another bound comes from the studies of galaxy
clusters, which gives $m_g < 2 \times 10^{-62}$g \cite{gold1974}.
Although this second limit is more restrictive, it is considered
less robust due to uncertainties in the content of the Universe in
large scales. Studying rotation curves of galactic disks, de
Araujo and Miranda \cite{deAraujo2007} have found that $m_g \ll
10^{-59}g$ in order to obtain a galactic disk with a scale length
of $b\sim 10$ kpc.

Studying the mass of the graviton in the weak field regime Finn
and Sutton have shown that the emission of gravitational radiation
does not exclude a non null (although small) rest mass. They found
the limit $m_g < 1.4 \times 10^{-52}$g \cite{Finn2002a} analyzing
the data from the orbital decay of the binary pulsars PSR B1913+16
(Hulse-Taylor pulsar) and PSR B1534+12.

In particular, as discussed by Bessada and Miranda
\cite{Bessada09}, if $m_g
> 10^{-65} g$ then massive gravitons would leave a clear signature
on the lower multipoles ($l< 30$) in the cosmic microwave
background (CMB) anisotropy power spectrum. Moreover, massive
gravitons give rise to a non-trivial Sachs-Wolfe effect which
leaves a vector signature of the quadrupolar form on the CMB
polarization \cite{Bessada2009}.

An interesting result that comes from Visser's model is that the
gravitational waves can present up to six polarization modes
\cite{paula2004} instead of the two usual polarizations obtained
from the GR. So, if in the future we would be able to identify the
gravitational wave polarizations, we would impose limits on the
graviton mass by this way.

The Visser's theory of massive gravitons can be used to build
realistic cosmological models that can be tested against available
observational data. It has the advantage that it is not necessary
to introduce new degrees of freedom neither extra cosmological
parameters. In fact, the cosmology with massive gravitons based on
the Visser's theory has the same number of parameters of the flat
$\Lambda$CDM model but no extra fields are added. In this paper we
derive cosmological constraints on the parameters of the Visser's
model. We use the most recent compilation of Type-Ia Supernovae
(SNe Ia) data, the so-called Union2 compilation of 557 SNe Ia
\cite{Amanullah2010}. We also combine the supernova data with
constraints from Baryon Acoustic Oscillations ({\rm BAO})
\cite{ApJ633_560} and {\rm CMB} shift parameter measurements
\cite{ApJS170_377}.

The paper is organized as follows: in Section \ref{sec:two} we
briefly review the Visser's approach. Section \ref{sec:three} is
devoted to the description of the cosmological model. In Section
\ref{sec:four} we investigate the observational constraints on the
Visser's cosmological model from SNe Ia, BAO and CMB shift
parameter data. In Section \ref{sec:five} we present our
conclusions.


\section{\label{sec:two}The Field Equations}


The full action considered by Visser is given by \cite{vis1998}:
\begin{eqnarray}
\label{fullaction}
I = \int d^4x\left[ \sqrt{-g}\frac{c^4R(g)}{16\pi G}
+ {\cal{L}}_{mass}(g,g_0) +{\cal{L}}_{matter}(g)\right],
\end{eqnarray}
where besides the Einstein-Hilbert Lagrangian and the Lagrangian
of the matter fields we have the bimetric Lagrangian
\begin{eqnarray}
{\cal{L}}_{mass}(g,g_0) = \frac{1}{2}m^2
\sqrt{-g_0}\bigg\{ ( g_0^{-1})^{\mu\nu}
( g-g_0)_{\mu\sigma}( g_0^{-1})^{\sigma\rho}
\\ \nonumber
 \times ( g-g_0)_{\rho\nu}-\frac{1}{2}
\left[( g_0^{-1})^{\mu\nu}( g-g_0)_{\mu\nu}\right]^2\bigg\},
\end{eqnarray}
where $m = m_g c/\hbar$, $m_g$ is the graviton mass and
$(g_0)_{\mu\nu}$ is a general flat metric.

The field equations, which are obtained by variation of
(\ref{fullaction}), can be written as:
\begin{equation}\label{field-equations}
G^{\mu\nu} -\frac{1}{2}m^2 M^{\mu\nu} = -\frac{8\pi G}{c^4}  T^{\mu\nu},
\end{equation}
where $G^{\mu\nu}$ is the Einstein tensor, $T^{\mu\nu}$ is the
energy-momentum tensor for perfect fluid, and the contribution of
the massive tensor to the field equations reads:
\begin{eqnarray}\label{massive tensor}
M^{\mu\nu} =  (g_0^{-1})^{\mu\sigma}\bigg[ (g-g_0)_{\sigma\rho}
- \frac{1}{2}(g_0)_{\sigma\rho}(g_0^{-1})^{\alpha\beta}\\
\nonumber
\times(g-g_0)_{\alpha\beta} \bigg](g_0^{-1})^{\rho\nu}    .
\end{eqnarray}

Note that if one takes the limit $m_g\rightarrow 0$ the usual
Einstein field equations are recovered.

Regarding the energy-momentum conservation we will follow the same
approach of \cite{Narlikar1984} and \cite{Rastall1972} in such a
way that the conservation equation now reads
\cite{Alves2006,Alves2009b}:
 \begin{equation}\label{conservation}
   \nabla_\nu T^{\mu\nu} = \frac{m^2 c^4}{16\pi G \hbar^2} \nabla_\nu M^{\mu\nu},
 \end{equation}
since the Einstein tensor satisfies the Bianchi identities
$\nabla_\nu G^{\mu\nu} = 0$.

\section{\label{sec:three}Cosmology with massive gravitons}
For convention we use the Robertson-Walker metric as the dynamical
metric:
\begin{equation}\label{rwmetric}
ds^2=c^2dt^2-a^2(t)\left[ \frac{dr^2}{1-kr^2}+r^2(d\theta^2+\sin^2\theta d\phi^2)\right],
\end{equation}
where $a(t)$ is the scale factor. The flat metric is written in
spherical polar coordinates:
\begin{equation}\label{minkmetric}
ds_0^2=c^2dt^2 - \left[dr^2 + r^2\left(d\theta^2 + \sin^2\theta d\phi^2  \right)   \right].
\end{equation}

The choice of Minkowski as the non-dynamical background metric
$g_0$ is based on the criterion of simplicity. In first place, the
metric $g_0$ is defined in such a way that it coincides with the
dynamical metric $g$ in the absence of gravitational sources. The
other point is that we do not need additional parameters for the
cosmological model. The last important point is that considering
Minkowski for $g_0$ we obtain a consistent relation for the
energy-momentum conservation law \cite{Alves2006}.

Using (\ref{rwmetric}) and (\ref{minkmetric}) in the field
equations (\ref{field-equations}) we get the following equations
describing the dynamics of the scale factor (taking $k=0$ for
simplicity):
 \begin{equation}\label{eqfried1}
   \left( \frac{\dot{a}}{a}\right)^2 + \frac{1}{4}m^2c^2(a^2 - 1)
= \frac{8\pi G}{3c^2} \rho
 \end{equation}
and
 \begin{equation}\label{eqfried2}
   \frac{\ddot{a}}{a}+\frac{1}{2}\left( \frac{\dot{a}}{a}\right)^2 + \frac{1}{8}m^2c^2a^2(a^2-1) = -\frac{4\pi G}{c^2}  p  ,
 \end{equation}
where as usual $\rho$ is the energy density and $p$ is the
pressure.

From Eq. (\ref{conservation}) we get the evolution equation for
the cosmological fluid, namely:
 \begin{equation}
   \dot{\rho} + 3 H \left[  (\rho + p) + \frac{m^2c^4}{32\pi G} (a^4 - 6a^2 + 3) \right] = 0,
 \end{equation}
where $H = \dot{a}/a$. Considering a matter dominated universe ($p
= 0$) the above equation gives the following evolution for the
energy density:
\begin{equation}\label{rho-m-new}
\rho = \frac{\rho_0}{a^3}- \frac{3m^2 c^4}{32\pi G}
\left( \frac{a^4}{7} - \frac{6a^2}{5} + 1 \right),
\end{equation}
where $\rho_0$ is the present value of the energy density. Note
that in the case $m_g \rightarrow 0 $ we obtain the usual
Friedmann equations.

Now, inserting (\ref{rho-m-new}) in the modified Friedmann
equation (\ref{eqfried1}) we obtain the Hubble parameter:
 \begin{equation}\label{parHubMass}
   H^2(a)=H^2_0\left[\frac{\Omega^0_m}{a^3}  + \frac{1}{2}\Omega_g^0 \left( 7a^2 -5a^4 \right)   \right],
 \end{equation}
where the relative energy density of the $i$-component is
$\Omega_i=\rho_i/\rho_c$ ($\rho_c=3H^2c^2/8\pi G$ is the critical
density) where `$i$' applies for baryonic and dark matter.
Moreover, the present contribution of the massive term is defined
by:
  \begin{equation}
    \Omega_g^0 = \frac{1}{70}\left( \frac{m_g}{m_H} \right)^2
  \end{equation}
where $m_H=\hbar H_0/c^2$ is a constant with units of mass.

Since we are assuming a plane Universe ($k=0$), the total density
parameter is $\Omega^0_{total} = 1$. Thus, $\Omega^0_g$ can be
replaced by $\Omega^0_g = 1-\Omega^0_m$. This tell us that the
model described by the Hubble parameter (\ref{parHubMass}) has
only two free parameters, namely $H_0$ and $\Omega^0_m$, which can
be adjusted by the cosmological observations, i.e., the same
number of free parameters of the $\Lambda$CDM model.


\section{Analysis and discussion}\label{sec:four}



\subsection{Supernova Ia}


In order to put constraints on the cosmological model derived from
the Visser's approach, we minimize the $\chi^2$ function
\begin{equation}
\label{chi2}
\chi^2(\Omega_m) = \sum_{i}\frac{\left[\mu_{th}(z_i|\Omega_m)
- \mu_{obs}(z_i)\right]^{2}}{\sigma^{2}(z_i)}
\end{equation}
where $\mu_{th}(z_i|\Omega_m)$ is the predicted distance modulus
for a supernova at redshift $z_i$. For a given $\Omega_m$ we have
\begin{equation}\label{distance modulus}
\mu(z|\Omega_m) \equiv m - M = 25 + 5\log d_L(z|\Omega_m),
\end{equation}
where $m$ and $M$ are, respectively, the apparent and absolute
magnitudes, and $d_L(z|\Omega_m)$ stands for the luminosity
distance given by
\begin{equation}
\label{luminosity dist}
d_L(z|\Omega_m)=(1+z)c\int^z_0
\frac{dz^{\prime}}{H(z^{\prime}|\Omega_m)}.
\end{equation}
Also, $\mu_{obs}(z_i)$ are the values of the observed distance
modulus obtained from the data and $\sigma(z_i)$ is the
uncertainty for each of the determined magnitudes from supernova
data.

Evaluating the minimum value of $\chi^2$ from the Union2
compilation of SNe Ia \cite{Amanullah2010} we found $\chi^2_{min}
= 561.11$ for the Visser's theory, with $\Omega_m =
0.261^{+0.021}_{-0.020}$, where we have considered errors at 1
sigma level.


\subsection{Baryon Acoustic Oscilations}


The primordial baryon-photon acoustic oscillations leave a
signature in the correlation function of luminous red-galaxies as
observed by Eisenstein et al. \cite{ApJ633_560}. This signature
provides us with a standard ruler which can be used to constrain
the following quantity
\begin{equation}
A = \sqrt{\Omega_m}E(z_1)^{-1/3}\left[ \frac{1}{z_1} \int_0^{z_1} \frac{dz}{E(z)} \right]^{2/3},
\end{equation}
where $E(z) = H(z)/H_0$, the observed value of $A$ is $A_{obs} =
0.469 \pm 0.017$ and $z_1 = 0.35$ is the typical redshift of the
SDSS sample. The computation of the value of $\Omega_m$ which
better adjust $A_{obs}$ lead us to $\Omega_m =
0.306^{+0.027}_{-0.025}$.


\subsection{CMB Shift Parameter}


The shift parameter R, which relates the angular diameter distance
to the last scattering surface with the angular scale of the first
acoustic peak in the CMB power spectrum, is given by (for $k=0$)
\cite{MNRAS291_L33, ApJS170_377}
\begin{equation}
R_{1089} = \sqrt{\Omega_{m}H_{0}^{2}}\int\limits_{0}^{1089}
\frac{dz}{H} = 1.70\pm 0.03.
\end{equation}
It is worth stressing that the measured value of $R_{1089}$ is
model independent. Also, note that in order to include the CMB
shift parameter into the analysis, it is needed to integrate up to
the matter-radiation decoupling ($z \simeq 1089$), so that
radiation is no longer negligible and it was properly taken into
account. With these considerations, the best-fit value for the
relative matter density using $R_{1089}$ is $\Omega_m =
0.224^{+0.046}_{-0.038}$.

\begin{table*}
\centering
\begin{tabular}{ccccc}
\hline
\hline
    & \multicolumn{2}{c}{Visser}               &              \multicolumn{2}{c}{$\Lambda$CDM} \\ \hline
Fit & $\chi^2_{min}$        & $\Omega_m$                   & $\chi^2_{min}$     & $\Omega_m$                \\ \hline
\\
SNe                 & $561.11$ & $0.261^{+0.021}_{-0.020}$ & $542.68$ & $0.270^{+0.021}_{-0.020}$ \\
CMB                 & $\sim 0$ & $0.224^{+0.046}_{-0.038}$ & $\sim 0$ & $0.239^{+0.043}_{-0.036}$ \\
BAO                 & $\sim 0$ & $0.306^{+0.027}_{-0.025}$ & $\sim 0$ & $0.273^{+0.025}_{-0.024}$ \\
SNe$+$CMB$+$BAO     & $565.06$ & $0.273^{+0.015}_{-0.015}$ & $543.76$ & $0.267^{+0.015}_{-0.015}$ \\
SNe(Sys)            & $538.83$ & $0.295^{+0.039}_{-0.036}$ & $530.72$ & $0.275^{+0.040}_{-0.037}$ \\
SNe(Sys)$+$CMB$+$BAO&$542.07$  & $0.290^{+0.020}_{-0.019}$ & $531.81$ & $0.265^{+0.019}_{-0.018}$ \\
\\ \hline
\end{tabular}
\caption{Best-fit values for $\Omega_m$ for the cosmological
observables considered in this work. It is also shown how the
introduction of systematic errors from the SNe measurements can
affect the best-fit. We have worked only with flat Universe
models, i.e., $k = 0$.} \label{results}
\end{table*}

\begin{figure}
\center{
\includegraphics[width=80mm]{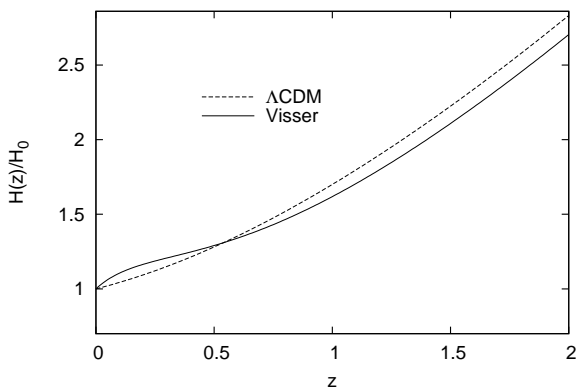}
}
\caption{Hubble parameter as a function of the redshift for
best-fit value obtained from SNe Ia. By using the different best-fit values, the curve does not change significatively}
\label{hubble}
\end{figure}

\begin{figure}
\center{
  \includegraphics[width=90mm]{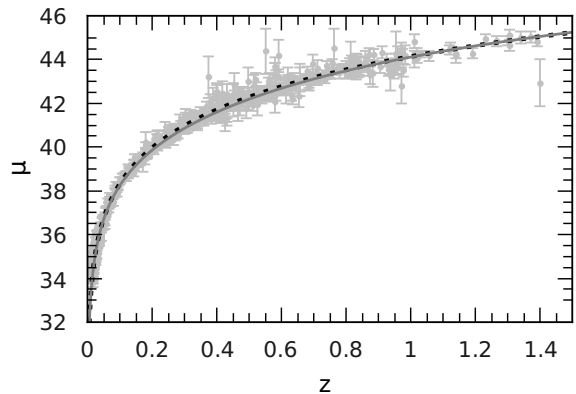}
}
\caption{Best-fit for the distance modulus versus redshift for the Visser model (solid gray line)
and the $\Lambda$CDM model (dashed line). The SNe data were taken from the Union2 compilation \cite{Amanullah2010}.}
\label{view}
\end{figure}


\subsection{Joint analysis}


When the measurements of SNe Ia luminosity distances are combined
with information related to the Baryon Acoustic Oscillation (BAO)
peak and the CMB shift parameter, the constraining power of the
fit to the parameters in the cosmological model is greatly
improved. Following such an approach we examine here the effects
of summing up the contributions of these last two parameters into
the $\chi^{2}$ of Eq. (\ref{chi2}). Our result is $\Omega_m =
0.273\pm 0.015$ with the corresponding minimum value for the
$\chi^2$ function: $\chi^2_{min} = 565.06$.

We can compare our results with the $\Lambda$CDM model by taking
the difference between $\chi^2_g$ and $\chi^2_{\Lambda CDM}$,
which are the minimum $\chi^2$ values for the massive bimetric
model and for the $\Lambda$CDM model, respectively. The evaluation
of this difference gives the result $\Delta \chi^2 = \chi^2_g -
\chi^2_{\Lambda CDM} = 21.30$, which shows that the bimetric
Visser's model is disfavored when compared with the flat
$\Lambda$CDM model.

In the Table \ref{results} we summarize our results for $\Omega_m$
considering each cosmological observable: SNe, CMB, BAO and the
combined analysis (SNe+CMB+BAO). For the sake of comparison it is
also shown the values of $\chi^2_{min}$ and $\Omega_m$ for the
$\Lambda$CDM model.

It is also instructive to evaluate the effect of adding the
systematic uncertainties of the SNe analysis on our results.
Considering only SNe, the addition of the systematic erros to the
statistical erros lead us to $\Omega_m = 0.295^{+0.039}_{-0.036}$
for the Visser's model. We also obtain a considerable lower value
for the difference between the $\chi^2$ of the two models $\Delta
\chi^2 = 8.11$. Now, taking into account the CMB and BAO
measurements together with SNe, we obtain $\Omega_m =
0.290^{+0.020}_{-0.019}$ and $\Delta \chi^2 = 10.26$ (see Table
\ref{results}).

In the Fig. \ref{hubble} and Fig. \ref{view} we show the Hubble
parameter and the distance modulus as functions of redshift
considering the best-fit value of $\Omega_m$ for the SNe. For the
sake of comparison, the standard $\Lambda$CDM model is also shown.
Note that although the massive graviton model is disfavored, it
seems to be able to reproduce very well the SNe Ia measurements,
as can be seen in the Fig. \ref{view}. This shows the importance
of the $\chi^2$ test in distinguishing the two models.


\subsection{Effective equation of state}

\begin{figure}
\center{
  \includegraphics[width=80mm]{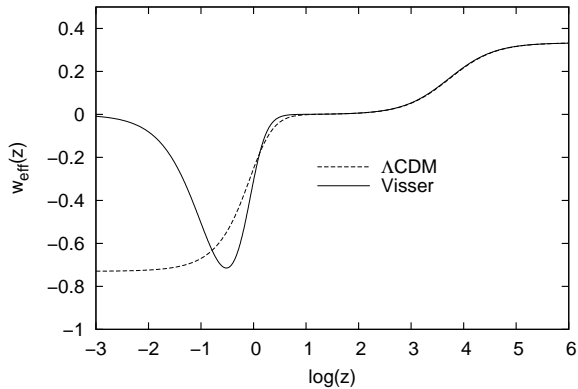}
}
\caption{Effective state parameter as a function of the redshift for the
best-fit value obtained from SNe Ia.}
\label{results1a}
\end{figure}
\begin{figure}
\center{
  \includegraphics[width=80mm]{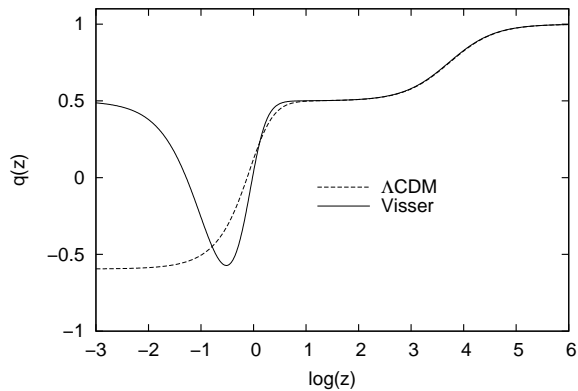}
}
\caption{Deceleration parameter as a function of the redshift for the
best-fit value obtained from SNe Ia. Note that the acceleration phase is transient in the Visser model.}
\label{results1b}
\end{figure}

The Fig. (\ref{results1a}) shows the effective equation of state
\begin{equation}
\label{eeos}
w_{eff}(z)=-1 +\frac{2(1+z)}{3H}\frac{dH}{dz}
\end{equation}
as a function of the redshift for the best-fit values above. The
deceleration parameter, which is shown in the Fig.\ref{results1b},
is related to $w_{eff}$ through $q(z) = (3w_{eff}(z) + 1)/2$. In
order to plot these curves we have included a component of
radiation with the present value of the density parameter
$\Omega_r= 5\times 10^{-5}$. For the best-fit value found in our
analysis, the Visser model goes through the last three phases of
cosmological evolution, i.e., radiation-dominated $(w = 1/3)$,
matter-dominated $(w = 0)$ and the late time acceleration phase
$(w < -1/3)$.

Note that for low redshifts the Visser's model shows additionally
a phase dominated by matter, indicating that for this model the
late time acceleration of the Universe was a transient phase which
has already finished. Moreover, for low redshifts, this behavior
of the Visser's theory is in accordance with the fact that the
luminosity distance of very low redshift SNe Ia can be fitted with
CDM model only, i.e., at very low redshift the $\Lambda$CDM, CDM
and Visser's model are degenerate for the cosmological
observations.

\section{Conclusions}\label{sec:five}


The theory of massive gravitons as considered in the Visser's
approach has the advantage that the field equations
(\ref{field-equations}) differs from Einstein equations only in a
subtle way, namely, by the introduction of the bimetric mass
tensor $M_{\mu\nu}$. Moreover the van Dam-Veltmann-Zakharov
discontinuity (vDVZ) present in the Pauli-Fierz term can be
circumvented in Visser's model by introducing a non-dynamical
flat-background metric \cite{will2006}.

From the cosmological point of view, the meaning of the mass
tensor, classically speaking, is a long range correction to the
ordinary Friedmann equation. Such a correction mimics the effects
of a dark energy component in such a way that additional fields
are not necessary.

In this context, we have shown that the cosmological model with
massive gravitons could be a viable explanation to the dark energy
problem. But, although the parameter $\Omega_m$ is well
constrained, the model is disfavored when compared to the
$\Lambda$CDM model. Considering systematic errors, the difference
between the $\chi^2_{min}$ of the two models reduces considerably,
but the the Visser model is still disfavored.

Finally, the plots of the effective state parameter and of the
deceleration parameter for the best fit value of $\Omega_m$, show
a very particular feature of the Visser's model, namely, the
transient behavior of the accelerated phase of expansion. The
Universe begins to accelerate approximately at the same redshift
of the $\Lambda$CDM model, but for a very small redshift ($z \sim
4 \times 10^{-2}$) we have a second transition and the Universe
becomes to decelerate again. In spite of this, the behavior of the
Hubble parameter $H(z)$ is very similar in both models as can be
seen in the Fig.\ref{hubble}. In this way, one would think that
the transient acceleration phase is what make the Visser model
less compatible with SNe data than the $\Lambda$CDM model. This is
a problem which we will address in the future in order to find
consistent modifications of Visser's approach.

\section*{Acknowledgments}
MESA would like to thank the Brazilian Agency FAPESP for support
(grant 06/03158-0). ODM and JCNA would like to thank the Brazilian
agency CNPq for partial support (grants 305456/2006-7 and
303868/2004-0 respectively). CAW thanks CNPq for support through
the grant 310410/2007-0. FCC acknowledges the postdoctoral
fellowship from FAPESP number 07/08560-4. The authors would like
to thank the referee for helpful comments that we feel
considerably improved the paper.

\newpage 
\bibliography{visserSNIa_bib}

\end{document}